\input{aipcheck}

\documentclass[
    ,final            
  ]
  {aipproc}

\layoutstyle{6x9}
\def\d{\mathrm{d}}

\begin{document}

\title{Diffractive $\rho$ production with an AdS/QCD holographic wavefunction for the $\rho$ meson}


\author{Jeff Forshaw}{
  address={University of Manchester, Oxford Road, Manchester M13 9PL, UK.}
}

\author{Ruben Sandapen}{
  address={Universit\'e de Moncton, Moncton, N-B, E1A 3E9, Canada \\ \& Mount Allison University, Sackville, N-B, E46 1E6, Canada.}
  }

\begin{abstract}
We report on the results of our recent research published in \cite{Forshaw:2012im}  that shows that AdS/QCD generates predictions for the rate of diffractive $\rho$-meson
electroproduction that are in agreement with data collected
at the HERA electron-proton collider \cite{Chekanov:2007zr,Collaboration:2009xp}. Preliminary results of this research were presented in \cite{Forshaw:2012iu}.  
 \end{abstract}

\maketitle


\section{Introduction}
We demonstrate another
success of the AdS/QCD correspondence \cite{Erdmenger:2007cm,CasalderreySolana:2011us,Costa:2012fw,deTeramond:2012rt} by showing that parameter-free AdS/QCD wavefunctions for the $\rho$
meson lead to predictions for the rate of diffractive $\rho$ meson
production ($\gamma^* p \to \rho p$) that agree with
the data collected at the HERA $ep$ collider. To compute the rate for diffractive $\rho$ production, we use the dipole model of high-energy scattering \cite{Nikolaev:1990ja,Nikolaev:1991et,Mueller:1993rr,Mueller:1994jq} in which the scattering amplitude for diffractive $\rho$ meson production is a convolution of the photon
and vector meson $q\bar{q}$ light-front wavefunctions with the total cross-section
to scatter a $q\bar{q}$ dipole off a proton. QED is used to determine
the photon wavefunction and the dipole cross-section can be extracted from the precise data on the deep-inelastic structure
function $F_2$ \cite{Soyez:2007kg,Forshaw:2004vv}. This formalism can then be used to predict rates for vector meson production and diffractive DIS \cite{Forshaw:2003ki,Forshaw:2006np} or to
to extract information on the $\rho$ meson wavefunction using the HERA data on diffractive $\rho$ production \cite{Forshaw:2010py,Forshaw:2011yj}. Here we use it to predict the cross-sections for diffractive $\rho$ production using an AdS/QCD holographic wavefunction for the $\rho$ meson proposed by Brodsky and de T\'eramond \cite{deTeramond:2008ht}. With this AdS/QCD wavefunction, we also predict the second moment of the twist-$2$ Distribution Amplitude of the longitudinally polarized $\rho$ meson. We find good agreement with Sum Rules and lattice predictions.

\section{An AdS/QCD holographic wavefunction}
 The AdS/QCD wavefunction is given by \cite{Forshaw:2012im,deTeramond:2008ht}
\begin{equation}
 \phi(x,\zeta)= N \frac{\kappa}{\sqrt{\pi}}\sqrt{x(1-x)} \exp \left(-\frac{\kappa^2 \zeta^2}{2}\right) \exp\left(-\frac{m_f^2}{2\kappa^2 x (1-x)} \right)
\label{lcwf-massive-quarks}
\end{equation}
where $N$ is a normalization constant and $m_f$ is the light quark mass.\footnote{Here we shall take $m_f = 140$~MeV, which is the value used in the dipole  fits to the deep-inelastic structure function, $F_2(x,Q^2)$ \cite{Forshaw:2011yj}.} In Eq. \eqref{lcwf-massive-quarks}, $\zeta=\sqrt{x(1-x)} b$ where $x$ is the light-front longitudinal momentum fraction of the quark and $b$ is the transverse separation between the quark and antiquark. $\zeta$ is the variable that maps onto the coordinate $z$ in the fifth dimension of AdS space \cite{deTeramond:2008ht}. The parameter $\kappa=M_{\rho}/\sqrt{2}=0.55$ GeV \cite{Forshaw:2012im}.  

This AdS/QCD wavefunction is rather similar to the original Boosted Gaussian (BG)
wavefunction discussed in \cite{Nemchik:1996cw,Forshaw:2003ki}
\begin{equation}
\phi^{{\mathrm{BG}}} (x,\zeta) \propto x(1-x) \;
\exp \left(\frac{m_f^{2}R^{2}}{2}\right)
\exp \left(-\frac{m_f^{2}R^{2}}{8 x(1-x)}\right) \; \exp \left(-\frac{2 \zeta^{2}}{{R}^{2}}\right) \;.
\label{original-boosted-gaussian} 
\end{equation}
If $R^2=4/\kappa^2$ then
the two wavefunctions differ only by a factor of $\sqrt{x(1-x)}$,
which is not surprising given that
in both cases confinement is modelled by a harmonic oscillator \cite{Forshaw:2012im}. In what follows we shall consider a parameterization that accommodates
both the AdS/QCD and the BG wavefunctions: 
\begin{equation}
 \phi(x,\zeta) \propto [x(1-x)]^\beta \exp \left(-\frac{\kappa^2 \zeta^2}{2}\right) \exp\left(-\frac{m_f^2}{2\kappa^2 x (1-x)} \right)~.
\label{lcwf-massive-quarks-fit}
\end{equation}
The AdS/QCD wavefunction is obtained by fixing $\beta=0.5$ and $\kappa=0.55$ GeV where as the BG wavefunction is obtained by fixing $\beta=1$ and treating $\kappa$ as a free parameter. 

The meson's light-front wavefunctions can be written in terms of the wavefunction $\phi(x,\zeta)$ \cite{Forshaw:2011yj}. For longitudinally
polarized mesons,
\begin{equation}
\Psi^{L}_{h,\bar{h}}(b,x) = \frac{1}{2\sqrt{2}}
\delta_{h,-\bar{h}} 
\left( 1 +  \frac{ m_{f}^{2} -  \nabla^{2}}{M_{\rho}^2\; x(1-x)}\right) \phi_L(x,\zeta) ~,
\label{nnpz_L}
\end{equation}
where $\nabla^2 \equiv \frac{1}{b} \partial_b + \partial^2_b$ and
$h$ ($\bar{h}$) are the helicities of the quark (anti-quark) and for  transversely polarized mesons, 
\begin{equation}
\Psi^{T=\pm}_{h,\bar{h}}(b, x) = \pm [i e^{\pm i\theta} 
( x \delta_{h\pm,\bar{h}\mp} - (1-x) \delta_{h\mp,\bar{h}\pm}) 
\partial_{b}+ m_{f}\delta_{h\pm,\bar{h}\pm}] \frac{\phi_T(x,\zeta)}{2x(1-x)}~,
\label{nnpz_T}
\end{equation}
where $be^{i\theta}$ is the complex form of the transverse separation,
$\mathbf{b}$.
We impose the normalization condition: 
\begin{equation}
\sum_{h,\bar{h}}\int \d^{2}{\mathbf{b}} \, \d x  \,
|\Psi^{\lambda}_{h,\bar{h}}(b, x)|^{2} = 1 ~,
\label{normalisationTL}
\end{equation}
where $\lambda=L,T$. This means we allow for a polarization dependent
normalization (hence the subscripts on $\phi_{L,T}$). 

\section{Comparison to data, Sum Rules and the lattice}
We first compute the AdS/QCD prediction for the electronic decay width
$\Gamma_{e^+e^-}$, which is related to the decay constant via
$$ f_\rho  = \left(  \frac{3\Gamma_{e^{+}e^{-}} M_\rho}{4 \pi
    \alpha_{\mathrm{em}}^2} \right)^{1/2}.$$
Using 
\begin{equation}
 f_\rho  = \frac{1}{2}
\left(\frac{N_c}{\pi}\right)^{1/2} \int_0^1 \d x \left(1 + \frac{m_{f}^{2}-\nabla^{2}}{M_\rho^2x(1-x)}\right) \phi_L(x,\zeta=0)~, 
\label{longdecayB}
\end{equation}
we obtain $\Gamma_{e^{+}e^{-}}=6.66$~keV, which is to be compared with the measured value $\Gamma_{e^{+}e^{-}}=7.04 \pm 0.06~\mathrm{keV}$
\cite{Nakamura:2010zzi}.   

\begin{figure}
 \includegraphics[height=.3\textheight]{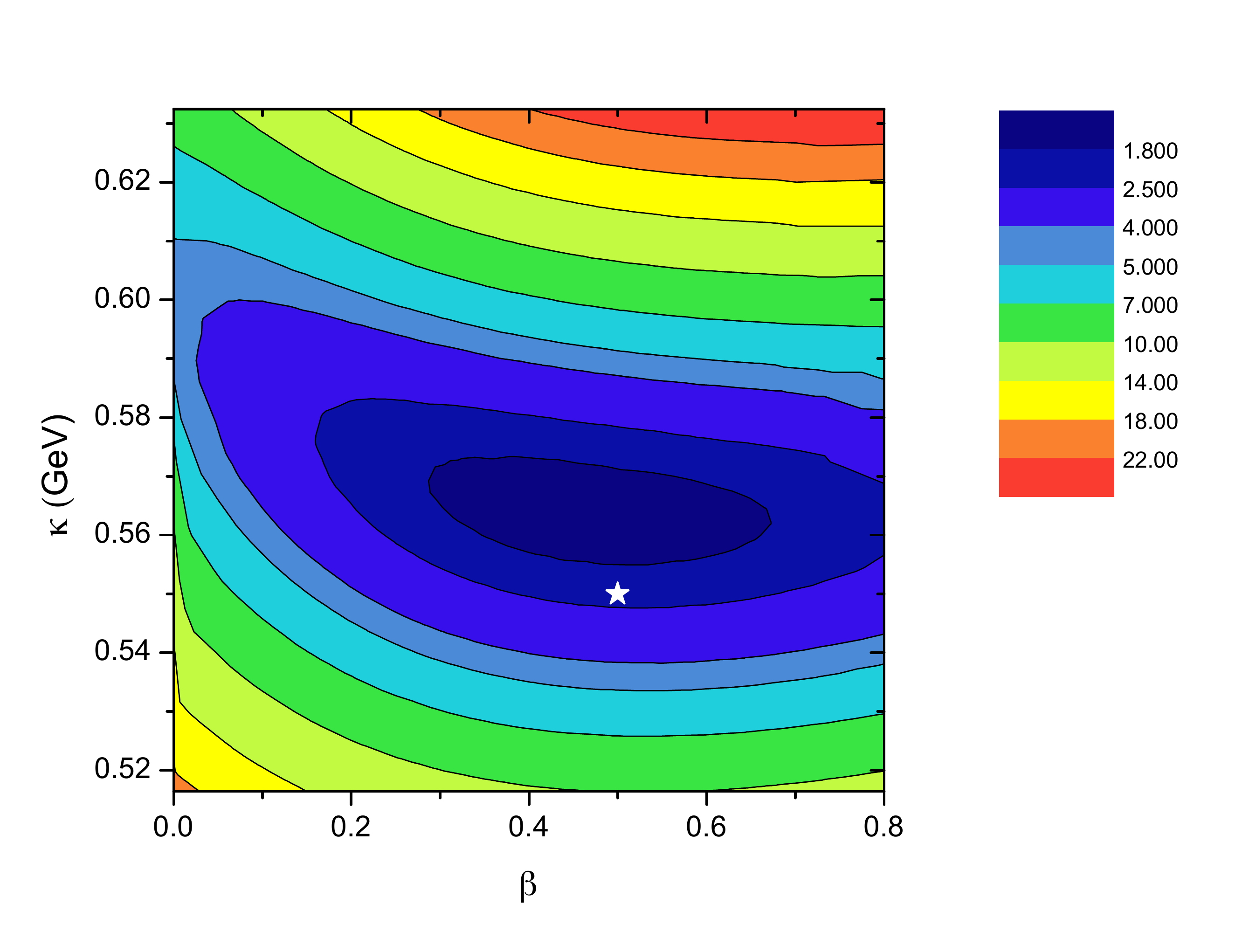}
  \caption{The $\chi^2$ distribution in the $(\beta,\kappa)$ parameter space.  The AdS/QCD prediction is the white star.}
\label{fig:contour}
\end{figure}

To  compute the rate for diffractive $\rho$ production, we use the CGC[0.74] dipole model \cite{Soyez:2007kg,Watt:2007nr}  (see \cite{Forshaw:2011yj} for details),
although the predicitions do not
vary much if we use other models that fit the HERA $F_2$ data
\cite{Forshaw:2004vv,Kowalski:2006hc}. The plots comparing the AdS/QCD predictions and the HERA data can be found in \cite{Forshaw:2012im}. Here, in figure \ref{fig:contour}, we show  the $\chi^2$ per data point\footnote{We include the
  electroproduction data and also the decay constant $f_\rho$ in the fit.}  in the
$(\beta,\kappa)$ parameter space (see
Eq. (\ref{lcwf-massive-quarks-fit})). It confirms that the AdS/QCD
prediction lies impressively close to the minimum in $\chi^2$. The best fit has a $\chi^2$ per data point equal to $114/76$ and is
achieved with $\kappa=0.56$ GeV and $\beta=0.47$. Note that the BG prediction i.e. $\beta=1, \forall \kappa$,  is clearly further away from the minimum in $\chi^2$.


We have previously shown \cite{Forshaw:2011yj} that the twist-$2$
Distribution Amplitude (DA) can be related to $\phi_L(x,\zeta)$
according to 
\begin{equation}
\varphi(x,\mu) = \left(\frac{N_c}{\pi}\right)^{1/2}  \frac{1}{2f_\rho}
\int \d
b \; \mu
J_1(\mu b) \left(1  + \frac{m_f^2 -\nabla^2}{M_\rho^2x(1-x)} \right) \phi_L(x,\zeta)~.
\label{tw2DAB}
\end{equation}
To compare to predictions using QCD Sum Rules \cite{Ball:2007zt} and from the lattice \cite{Boyle:2008nj},  we compute the moment:
\begin{equation}
 \int_0^1 \d x \;  (2x-1)^2 \varphi(x,\mu)~.
\end{equation}
We obtain a value of $0.228$ for the AdS/QCD wavefunction, which is to be compared with the
Sum Rule result of $0.24 \pm 0.02$ at $\mu = 3$~GeV \cite{Ball:2007zt} and the lattice
result of $0.24\pm 0.04$ at $\mu = 2$~GeV \cite{Boyle:2008nj}.  The AdS/QCD
wavefunction neglects the perturbatively known evolution with the
scale $\mu$ and should be viewed as a parametrization of the DA at
some low scale $\mu \sim 1$ GeV. Viewed as such, the agreement is
good.


\begin{theacknowledgments}
 RS thanks the organizing committee of Diffraction 2012 for their invitation, financial support and a very enjoyable conference. 
\end{theacknowledgments}



\bibliographystyle{aipproc}   

\bibliography{sandapen}

\IfFileExists{\jobname.bbl}{}
 {\typeout{}
  \typeout{******************************************}
  \typeout{** Please run "bibtex \jobname" to optain}
  \typeout{** the bibliography and then re-run LaTeX}
  \typeout{** twice to fix the references!}
  \typeout{******************************************}
  \typeout{}
 }

\end{document}